\documentclass[twocolumn,showpacs,preprintnumbers,aps,pra,amsmath,amssymb,floatfix]{revtex4}

\usepackage{graphicx}
\usepackage{dcolumn}
\usepackage{bm}

\begin{document}

\title{State-Selective Detection of Near-Dissociation Ultracold KRb
\\$X~^1\Sigma^+$ and $a~^3\Sigma^+$ Molecules}

\author{D. Wang}
\author{E. E. Eyler}
\author{P. L. Gould}
\author{W. C. Stwalley}
\affiliation{Physics Department, University of Connecticut, Storrs, CT 06269-3046}

\date{\today}

\begin{abstract}
We report on the state-selective detection of near-dissociation ultracold KRb molecules in 
the ground $X^1\Sigma^+$ state and the metastable $a^3\Sigma^+$ state. 
The molecules are 
produced by photoassociation of ultracold atoms followed by radiative 
decay into high vibrational levels of the $X$ and $a$ states. Detection utilizes 
resonance-enhanced one-color two-photon ionization, followed by time-of-flight
mass spectroscopy. Scanning the detection laser frequency over the range 582-625 
nm, we observe transitions from the $v''$=86-92 levels of the $X$ state, which 
are bound by up to 30 cm$^{-1}$, and the $v''$=17-23 levels of the $a$ state, which 
are also bound by up to 30 cm$^{-1}$. The measured vibrational spacings are in 
excellent agreement with those previously measured and those calculated from the
relevant potential curves. Relative
vibrational populations are also consistent with Franck-Condon factors for decay from
the photoassociated levels.
\end{abstract}



\maketitle

\section{\label{sec:level1}Introduction}

    Following in the footsteps of ultracold atoms, the area of ultracold molecules has 
developed rapidly in recent years \cite{Bahns00,Masnou01,Bethlem03}. Polar molecules, with their permanent 
electric dipole moments, have drawn particular attention \cite{EPJD04}. The dipole moments 
of these heteronuclear molecules allow manipulation with applied electric fields. 
Because the dipole-dipole potential is both long-range and anisotropic, interactions 
between polar molecules are fundamentally different from those in homonuclear 
systems. Potential applications include novel quantum degenerate systems \cite{Santos00,Yi00,Goral00,Goral02a,Damski03,Goral02b,Baranov02},
quantum computation \cite{DeMille02}, and tests of fundamental
symmetries \cite{Kozlov95}. Investigations of collisions and reactions at these extremely
low temperatures also promise to open new areas of ultracold chemistry \cite{Bohn01,Bodo02,Balakrishnan01}.

    The technique of ultracold atom photoassociation (PA) \cite{Stwalley99} has been the primary 
means for producing molecules at translational temperatures below 1 mK \cite{Masnou01}. 
In this process, laser light resonantly binds two colliding atoms into an excited 
molecule which subsequently decays by spontaneous emission. This method has recently 
been successfully adapted to heteronuclear systems \cite{Kerman04,Sage05,Bergeman04,Mancini04,Shaffer99,Haimberger04,Wang03,Wang04a,Wang04b}. Feshbach resonances have 
been used to make homonuclear molecules at even lower temperatures \cite{Herbig03,Chin03,
Durr04,Xu03,Jochim03,Cubizolles03,Strecker03,Zwierlein03,Regal03,Greiner03}, in some 
cases under quantum degenerate conditions. Based on recent observations of Feshbach 
resonances in heteronuclear systems \cite{STA04,Inouye04}, it can be expected that this route 
will also yield ultracold heteronuclear molecules in due course. Both PA and 
Feshbach resonances tend to produce molecules in high vibrational levels with 
outer turning points at long range. Although these high-$v''$ states are of interest 
for some studies, many applications require low-$v''$ states, because of their 
improved stability against inelastic processes and their larger dipole moments. 
Various schemes for populating low-$v''$ states have been proposed \cite{Band95,DeMille02,Bergeman04,Kotochigova04,Stwalley04,Damski03},
and some successful implementations have been reported \cite{Nikolov00,Sage05}.

    Direct detection of ultracold molecules has generally utilized photoionization 
combined with time-of-flight mass spectroscopy. State selectivity is essential to many 
applications and for diagnosing transfer of population. For example, measuring 
the $v''$ dependence of vibrational quenching rates due to collisions with ultracold 
atoms would be an important first step in studying ultracold molecule collisions. To date, 
the ability to unambiguously identify the vibrational state of ultracold molecules 
has been limited, particularly for levels near dissociation. Two-photon one-color
ionization spectra have been reported 
in Rb$_2$ \cite{Gabbanini00,Fioretti01,Kemmann04}, but definitive assignments and initial state 
identifications were not made. We have recently demonstrated state-selective detection 
of Rb$_2$, details of which will be reported elsewhere \cite{Huang05}. In Cs$_2$, ultracold
molecule detection spectra have been compared to absorption measurements, revealing the role
of the diffuse bands in the detection process \cite{Dion02}. In K$_2$, two-photon two-color
ionization spectra 
enabled some identification of the low-$v''$ states produced by both one-photon \cite{Nikolov99} 
and two-photon \cite{Nikolov00} PA. Dissociation with a separate cw laser helped to clarify 
the assignment \cite{Nikolov00}. In Na$_2$, high-resolution cw ionization spectroscopy yielded 
clear identification of the highest-lying bound and quasibound rovibrational-hyperfine 
states \cite{Fatemi02}.

  To date, the only heteronuclear system in which state selectivity has been 
achieved is RbCs. Using two-photon two-color ionization, ultracold molecules 
were detected in specific vibrational levels of the $a^3\Sigma^+$ state \cite{Kerman04,Bergeman04}. In 
addition, these molecules have been transferred by stimulated emission pumping 
to the $X^1\Sigma^+$ state ($v''$=0,1) and detected in these states with vibrational selectivity \cite{Sage05}.

    In the present work, we report on vibrationally state-selective detection of 
ultracold KRb molecules in high-$v''$ levels of both the ground $X^1\Sigma^+$ state and the 
metastable $a^3\Sigma^+$ state. These molecules are formed by cold-atom photoassociation 
followed by radiative decay, as shown in Fig. 1. Two-photon one-color ionization proceeds through 
resonant intermediate levels of the $4^1\Sigma^+$, $5^1\Sigma^+$, $4^3\Sigma^+$, and
$3^3\Pi$ states, allowing vibrational state identification and determination of the relative
populations. Using wavelengths in the range of 582-625 nm,
we have observed vibrational levels $v''$=86-92 for the $X$ state and $v''$=17-23
for the $a$ state. We have also analyzed spectroscopy of the excited states, details
of which will be reported elsewhere \cite{Wang05}.

\section{Experiment}

\begin{figure}
\centering
\includegraphics[width=0.95\linewidth]{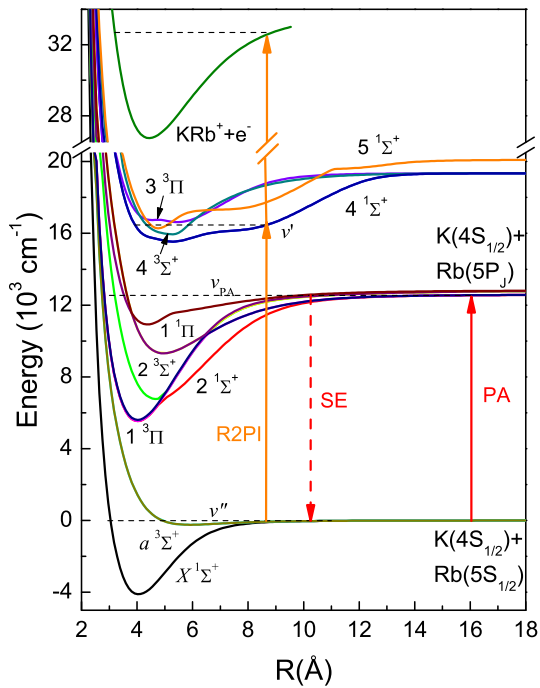}
 \caption{\protect\label{fig1} (Color online) Relevant KRb potential energy curves from \cite{Rousseau00}. A pair of colliding 
ground-state atoms is photoassociated (PA) to a bound level $v_{PA}$ below the K(4$S$)+Rb(5$P$) 
asymptote which subsequently undergoes spontaneous emission(SE) to either the $X^1\Sigma^+$
or the $a^3\Sigma^+$ state. These states are detected by resonant two-photon ionization (R2PI)
through the $4^1\Sigma^+$ and $5^1\Sigma^+$ states (for singlet molecules) or the $4^3\Sigma^+$
and $3^3\Pi$ states (for triplet molecules).}
\end{figure}

    Details of the experimental setup have been described previously \cite{Wang04a,Wang04b}. 
Here we recount it briefly, focusing on the ionization detection. 
The KRb PA takes place in overlapping clouds of ultracold $^{39}$K and $^{85}$Rb. 
High atomic densities, estimated at $3\times10^{10}$ cm$^{-3}$ for K and $1\times10^{11}$ cm$^{-3}$ 
for Rb, are achieved using ``dark-SPOT'' MOTs for each species. Temperatures for K and Rb 
of 300 $\mu$K and 100 $\mu$K, respectively, are expected.

    The PA process is driven with a cw tunable titanium-sapphire laser 
(Coherent 899-29). Its output, typically $>$400 mW, is focused into 
the overlapping MOT clouds. PA spectra, as described 
previously \cite{Wang04a,Wang04b}, are obtained by scanning this laser and 
measuring the ionization signal from molecules which have 
radiatively decayed into the $X^1\Sigma^+$ ground state and the metastable $a^3\Sigma^+$ state.

  In our earlier work \cite{Wang04a,Wang04b}, the ionization detection used a pulsed laser
with a broader linewidth and significant amplified spontaneous emission (ASE). This
prevented state-selective detection but had the benefit of yielding at least some ion
signal at most wavelengths, thereby facilitating the location of PA resonances.

  In the present work, ionization detection is achieved with a pulsed dye laser 
(Continuum ND6000) pumped by a frequency-doubled Nd:YAG 
laser at a 10 Hz repetition rate. The use of two dyes, R610 and DCM, provides spectral
coverage over the range 582 nm to 625 nm. The 0.05 cm$^{-1}$ linewidth of the dye laser
is sufficient to resolve the vibrational structure, but not the rotational structure,
of transitions 
from levels of the $X^1\Sigma^+$ and $a^3\Sigma^+$ states. This detection laser, with 
a pulse width of 7 ns and a typical output power of 3 mJ, is 
focused to a diameter of 1 mm. This is significantly larger 
than the 0.3 mm diameter of the MOT clouds in order to illuminate 
a larger fraction of the ballistically expanding cloud of cold 
molecules. Ions from the laser pulse are accelerated to a channeltron 
ion detector. KRb$^+$ is discriminated from other species 
(K$^+$, Rb$^+$, Rb$_2^+$) by its time of flight (TOF).
	
\section{Detection Spectra for Singlet Molecules}

    Detection spectra are obtained by fixing the PA frequency on a resonance 
and recording the KRb$^+$ ion signal while scanning the pulsed laser. A 400 cm$^{-1}$
scan for $X^1\Sigma^+$ state molecules is shown in Fig. 2. The spectra display
structure on two scales. On a gross scale
($\sim$20 cm$^{-1}$), nearly periodic spacings correspond to vibrational levels of the upper state 
of the detection transition, specifically the $4^1\Sigma^+$ state for the spectrum 
shown in Fig. 2. The spectroscopy of this upper state will be 
described separately \cite{Wang05}. Measured level spacings 
and the range of levels observed are both in very good agreement with 
calculations based on $ab$ $initio$ potentials \cite{Rousseau00}. We note that the increased
complexity of the spectrum at higher frequencies is likely due to the appearance of transitions
to the $5^1\Sigma^+$ state, which are not yet assigned.

\begin{figure}
\centering
\includegraphics[width=1.00\linewidth]{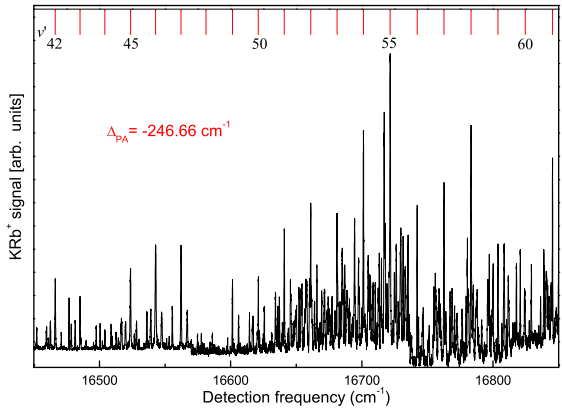}
 \caption{\protect\label{fig2} (Color online) Detection spectrum for singlet molecules. The KRb$^+$ ion signal is plotted 
as a function of the detection laser frequency. The PA laser is tuned to a $3(0^+)$ 
($J$=1) level at 12569.94 cm$^{-1}$, which corresponds to a detuning of $-$246.66 cm$^{-1}$ 
below the K(4$S$)+Rb(5$P_{3/2}$) asymptote. Each group of lines is labeled according 
to the $v'$ level of the $4^1\Sigma^+$ state which is excited. The strongest line in each
group corresponds to a transition from $v''$=89 to the $v'$ indicated. The blue end of the 
spectrum becomes congested and is not yet assigned.}
\end{figure}

\begin{figure}
\includegraphics[width=1.00\linewidth]{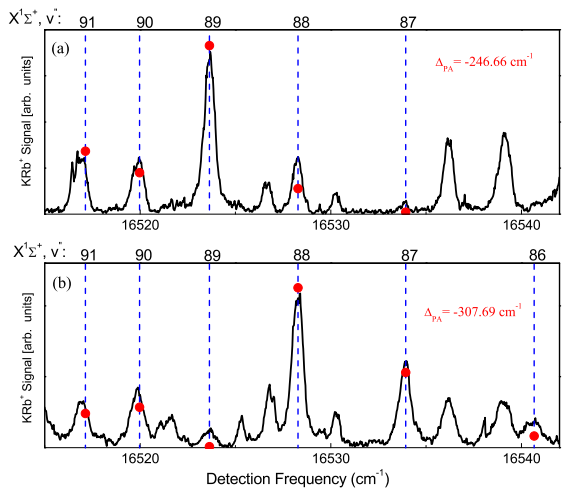}
 \caption{\protect\label{fig3} (Color online) An expanded view of the $v'$=45 group of lines in the singlet detection spectrum 
for two different $3(0^+)$ PA detunings: (a) $-$246.66 cm$^{-1}$ and (b) $-$307.69 cm$^{-1}$. 
The calculated $X$-state level spacings (dashed vertical lines) are superimposed. Most
unlabelled lines belong to adjacent $v'$ groups. Solid circles 
indicate the calculated FCF$'$s for decay from 3($0^+$) to $X^1\Sigma^+$ for these two cases. 
The maximum FCF value in (a) is 0.11 and the maximum in (b) is 0.12. Note the shift of 
the FCF peak to lower $v''$ (smaller outer turning point) with larger PA detuning.}
\end{figure}

    On a finer scale ($\sim$5 cm$^{-1}$), the structure corresponds to the 
spacings between high-lying vibrational levels of the $X^1\Sigma^+$ ground 
state. Specific $X$-state levels are assigned by matching the measured 
spacings with those calculated from the potential \cite{Zemke04}.
Examples are shown in Fig. 3 for PA to two different vibrational levels of 
the $3(0^+)$ state. As can be seen, the agreement between measured and 
predicted vibrational spacings is excellent, allowing unambiguous 
identification of the ground-state levels. A previously
measured spacing between $v''$=86 and 87 \cite{Amiot00} is also in excellent agreement.
We note that although the $3(0^+)$ state comes from the K(4$s$) + Rb(5$p_{3/2}$)
asymptote, the PA laser is tuned below the K(4$s$) + Rb(5$p_{1/2}$) asymptote, so
predissociation does not occur.

    The relative populations of the various $X$-state vibrational levels are 
determined by the Franck-Condon factors (FCF$'$s) for decay of the 
level formed by photoassociation. We expect that as PA occurs to more deeply 
bound levels, with smaller outer turning points, the FCF$'$s will favor 
decay to more deeply bound levels of the $X$-state. This is indeed the 
case, as seen in comparing Figs. 3(a) and 3(b). The peak of the $X$-state 
distribution shifts to a lower $v''$ for the larger (more negative) PA detuning. 
If we assume that the first (resonant) step of the two-photon detection 
process is saturated, and that the second (ionization) step is saturated 
and/or structureless, then the detection spectra will be a direct measure 
of the relative population of $X$-state vibrational levels. This distribution 
is given by the FCF$'$s for decay of the photoassociated level, in this case 
a level of the $3(0^+)$ state. These FCF$'$s, calculated using 
the LEVEL program \cite{LEV74}, are superimposed on the spectra 
in Figs. 3(a) and 3(b). Not only does the 
measured peak of the distribution shift with detuning as predicted, but 
the relative peak heights within each distribution actually match the 
FCF calculations rather well.

    Using PA detunings for the $3(0^+)$ state over the range \-246 cm$^{-1}$ to \-320 cm$^{-1}$ 
(measured relative to its K(4$S$) + Rb(5$P_{3/2}$) asymptote, which lies 237.60 cm$^{-1}$ above 
the K(4$S$) + Rb(5$P_{1/2}$) asymptote), we observe $X$-state levels from $v''$=86-92, 
which have binding energies from 29.76 cm$^{-1}$ to 4.45 cm$^{-1}$, respectively. 
The highest level in the $X$ state is predicted \cite{Zemke04} to be $v''$=98. The molecules 
we detect are produced over a time interval of several milliseconds 
before the detection laser pulse. During this time, they are exposed to the Rb MOT
trap (and repump) light, tuned near the K(4$S$) + Rb(5$P_{3/2}$) asymptote, and PA 
light. This could cause off-resonant reexcitation and subsequent 
dissociation of the $X$-state molecules, particularly for high $v''$. 
Such a state-dependent destruction mechanism would modify the $X$-state 
vibrational distribution from that intially produced according to the FCF$'$s. 
However, we see no evidence for this alteration.

\section{Detection Spectra for Triplet Molecules}

\begin{figure}
\centering
\includegraphics[width=1.00\linewidth]{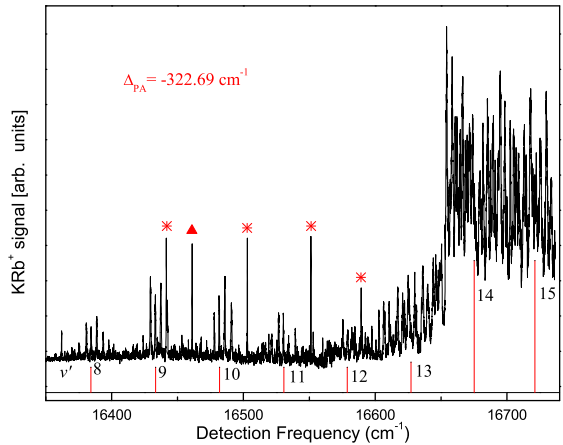}
 \caption{\protect\label{fig4} (Color online) Detection spectrum for triplet molecules. The PA laser is tuned to a 
3($0^-$) ($J$=1) level at 12493.91 cm$^{-1}$, which corresponds to a detuning of -322.69 
cm$^{-1}$ below the K(4$S$)+Rb(5$P_{3/2}$) asymptote. Each group of lines is labeled 
according to the $v'$ level of the $4^3\Sigma^+$ state which is excited. The blue end 
of the spectrum becomes congested due in part to the emergence of
transitions to the $3^3\Pi$ state, which have not yet been assigned. The triangle indicates
a one-photon atomic transition from Rb(5p$_{3/2}$)$\rightarrow$(5f). Asterisks indicate
two-photon transitions from Rb(5s) to atomic Rydberg states: 13d, 14d, 15d, 16d from left to right.}
\end{figure}

  State-selective detection of triplet $a^3\Sigma^+$ molecules is carried out in a similar 
manner to that described above for singlet $X^1\Sigma^+$ molecules. The only difference 
is that the PA laser is tuned to a level which decays
to the $a$ state. For this work, we use the 3($0^-$) state from the 
K(4$S$) + Rb(5$P_{3/2}$) asymptote, which correlates to the $1^3\Pi$ state at short
range \cite{Wang04b}. Higher
detection laser intensities are required for comparable signal 
levels from triplet molecules. The triplet detection spectra are obtained 
over the same wavelength range, but of course involve different upper states: $4^3\Sigma^+$ 
and probably $3^3\Pi$. The spectroscopy of the $4^3\Sigma^+$ 
state, along 
with that of the $4^1\Sigma^+$ state (used for singlet detection), will be described in a 
separate publication \cite{Wang05}. A $\sim$400 cm$^{-1}$ scan is shown in Fig. 4. The structure 
repeating at $\sim$50 cm$^{-1}$ corresponds to the vibrational spacing of the $4^3\Sigma^+$ 
state. Assigned vibrational levels are indicated in the figure. The congested region of larger
signal to the blue may involve the $3^3\Pi$ state, but specific assignments have not yet
been made.

  Two-photon transitions to atomic Rb Rydberg states sometimes can be seen as well
in the triplet spectra,
due to leakage from the space-charge-broadened Rb$^+$ TOF peak into the KRb$^+$ TOF peak. 
We also see ions at the Rb atomic 5p$_{3/2}\rightarrow$5f one-photon transition. The 5p$_{3/2}$
level is populated by the MOT lasers and this dipole-forbidden transition
is enabled by the $\sim$160 V/cm electric field used for extraction.
These atomic lines serve as useful frequency markers and verify that the laser 
frequency measurements are accurate to 0.16 cm$^{-1}$. These lines are not readily 
observable in the spectra of singlet molecules (Fig. 2) because lower laser 
intensities are used.
  
\begin{figure}
\centering
\includegraphics[width=1.00\linewidth]{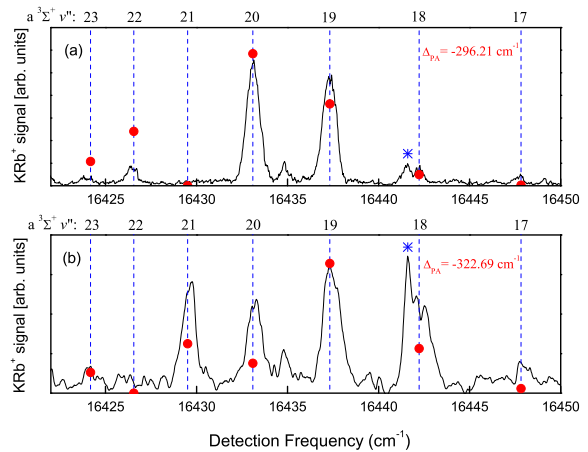}
 \caption{\protect\label{fig5} (Color online) An expanded view of the $v'$=9 group of lines in the
triplet detection
spectrum for two different 3($0^-$) PA detunings: (a) $-$296.21 cm$^{-1}$ and (b) $-$322.69 
cm$^{-1}$. The calculated $a$ state level spacings (dashed vertical lines) are 
superimposed. Solid circles indicate the calculated FCF$'$s for decay from the 
3($0^-$) state to the $a^3\Sigma^+$ state. The maximum FCF value in (a) is 0.10 ($v''=20$) and
the maximum in (b) is 0.10 ($v''=19$). Asterisks indicate the two-photon atomic transition:
Rb(5s)$\rightarrow$(13d)}.
\end{figure}

  Fig. 5 is an expanded view of the spectrum, encompassing only one vibrational 
level of the $4^3\Sigma^+$ upper state. The structure here is due to the near-dissociation 
levels of the $a^3\Sigma^+$ state. The vibrational levels show spacings of 2.4 to 5.6 cm$^{-1}$.
Such spacings correspond to levels $v''$=20 to 26 in the $ab$ $initio$ potential of
Kotochigova et al. \cite{Kotochigova03} and $v''$=18 to 24 in the $ab$ $initio$ potentials of
Park et al. \cite{Park00} and of Rousseau et al. \cite{Rousseau00}. These three potentials have
been compared by Zemke et al. \cite{Zemke05}. However, a definitive assignment has very recently
become available, based on new Fourier Transform Spectra in Hannover \cite{Tiemann05}. This clearly
indicates we have observed levels $v''$=17 to 23, as shown in Table 1. We also plan to calculate improved
Franck-Condon factors for the Hannover $a^3\Sigma^+$ potential once it is available. The FCFs in Fig. 5 are
based on the Kotochigova et al. potential \cite{Kotochigova03} for levels 20 to 26. Finally, we plan direct
measurement of the binding energies by scanning a separate cw laser to deplete the ground-state levels. We
have recently observed this ``ion dip'' spectroscopy for $X^1\Sigma^+$ $v''$=89.

Table I. Level spacings ($\Delta{G}_{v+1/2}$, in cm$^{-1}$) in the $^{39}$K$^{85}$Rb $a^3\Sigma^+$ state
\newline
\begin{tabular}{ c| c| c }

	\hline\hline
$v$ & Fourier Transform Spectra  & PA-REMPI(This~work)   \\
	\hline
17   & 5.49 & 5.60  \\
18   & 4.84 & 4.90  \\
19   & 4.23 & 4.23  \\
20   & 3.60 & 3.56  \\
21   & 2.99 & 2.96  \\
22   & 2.41 & 2.38  \\
	\hline\hline
\end{tabular}
\newline

  As for the singlet molecules, we can use the peak heights as a measure 
of the lower level ($a^3\Sigma^+$) populations. There is a small, but noticeable 
shift in the distribution to lower $v''$ for larger (more negative) PA detunings. 
Also shown in these figures are the calculated FCF$'$s for decay from each 3($0^-$) PA 
level to various levels of the $a^3\Sigma^+$ state. Calculations using 
the LEVEL program \cite{LEV74} reproduce the
overall locations of the distributions, including the shift with PA detuning. 
However, compared to the singlet spectra (Fig. 3), individual peak heights are 
not as accurately predicted. We hope future calculations based on the Hannover $a^3\Sigma^+$ potential,
once it is available, can give more accurate results. Linewidths in the triplet spectra are somewhat broader than
those in the singlet spectra. Power broadening, spin-spin, second-order spin-orbit, rotational and
hyperfine structure should all contribute to the line shapes observed.

  PA detunings for the 3($0^-$) state from $\-244$ cm$^{-1}$ to $\-$323 cm$^{-1}$
(measured relative to its K(4$P$) + Rb(5$P_{3/2}$) asymptote) have been used to
observe $a$-state levels from $v''$=17-23. These have binding energies from
29.02 cm$^{-1}$ to 5.31 cm$^{-1}$, respectively. This numbering, based on the
Fourier Transform Spectra in Hannover \cite{Tiemann05}, is definitive because vibrational
assignment is based on two different isotopes. None of the three sets of $ab$ $initio$ potentials from
\cite{Kotochigova03, Rousseau00, Park00} can give this exact numbering, although a good vibrational spacing match
can be found if we adjust their numbering by one or three \cite{Zemke05}.

\begin{figure}
\centering
\includegraphics[width=1.00\linewidth]{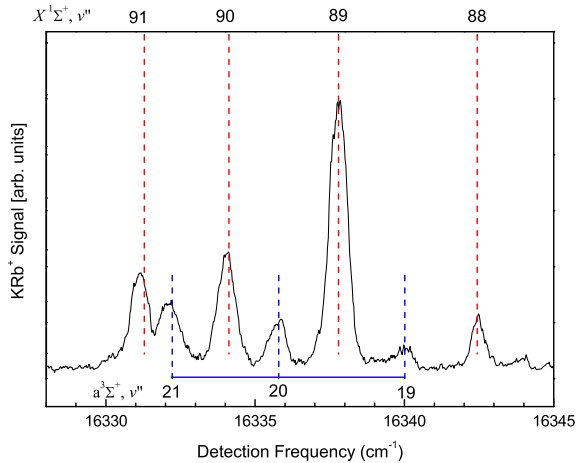}
 \caption{\protect\label{fig6} (Color online) Detection spectrum with overlapping singlet and triplet
features. The PA laser is tuned to a 3($0^+$) (J=1) level at a detuning of -246.66 cm$^{-1}$
below the K(4S) + Rb (5P$_{3/2}$) asymptote. $X^1\Sigma^+$ molecules are detected through
$4^1\Sigma^+$ ($v'$=33), while $a^3\Sigma^+$ molecules are detected though
$4^3\Sigma^+$ ($v'$=4). Expected singlet and triplet line positions are indicated.}
\end{figure}

  An important difference between singlet and triplet molecules is that the triplets 
have a non-zero magnetic moment and can therefore be magnetically trapped. We 
have previously demonstrated this trapping in the quadrupole magnetic field of 
the MOT \cite{Wang04a,Wang04b} by delaying the molecule detection with respect
to the turn-off of the PA laser. This difference in magnetic properties could be utilized 
as a singlet/triplet ``filter$"$ to distinguish the two types of molecules. We do
see cases where both $X^1\Sigma^+$ and $a^3\Sigma^+$ molecules appear in the
same region of the detection spectrum. An example, using PA to the 3($0^+$)
state, is shown in Fig. 6. At long range, the 3($0^+$) state should decay to
both $X^1\Sigma^+$ and $a^3\Sigma^+$ states \cite{Wang04b},
so we expect to detect both. However, the
FCF$'$s for the first step of the detection process play an important role.
For high-$v''$ levels of $X^1\Sigma^+$, the overlap with $4^1\Sigma^+$ levels comes primarily
from the outer turning point. On the other hand, overlap of high-$v''$ levels
of $a^3\Sigma^+$ with $4^3\Sigma^+$ levels comes primarily from the inner turning point and
becomes more favorable in the lower energy region of the spectrum. The
overall detection efficiencies (including the ionization step) for $X^1\Sigma^+$ and
$a^3\Sigma^+$ become comparable in the region shown in Fig. 6. Although we do see
triplet features in a primarily singlet detection spectrum (using PA to
3($0^+$)), we do not see singlet features in the triplet detection spectra
(using PA to 3($0^-$)). This is consistent with the fact that at long range,
3($0^+$) can decay to both to both $X^1\Sigma^+$ and $a^3\Sigma^+$ states,
while 3($0^-$) can decay only to $a^3\Sigma^+$ \cite{Wang04b}.

\section{Conclusions}

  In summary, we have realized vibrationally state-selective detection of near-dissociation 
levels of ultracold KRb molecules in the $X^1\Sigma^+$ ground state and the $a^3\Sigma^+$ 
metastable state. 
This state-selectivity will be crucial to future experiments in ultracold molecular 
collisions and reactions where specific initial and final states must be measured. 
This capability is equally important in diagnosing population transfer, e.g., from 
high-$v''$ to low-$v''$ \cite{Sage05}. In fact, the first step in our singlet detection 
($4^1\Sigma^+\leftarrow X^1\Sigma^+$) is a good candidate for realizing this type of transfer. 
As an 
example, if we start in $X^1\Sigma^+$ ($v''$=89), the FCF for excitation to 
$4^1\Sigma^+$ ($v'$=40) is 
quite large (0.02) due to overlap at the outer turning points. On the other hand, overlap 
at the inner turning points gives a favorable FCF of $\sim$0.01 for decay (or stimulated emission) 
of $4^1\Sigma^+$ ($v'$=40) to the absolute ground state, $X^1\Sigma^+$ ($v''$=0). 
These large FCF$'$s indicate 
that coherent two-photon transfer, such as STIRAP \cite{Bergman98}, should be feasible with
narrow-linewidth quasi-cw lasers.

  The two-photon, one-color detection we have employed is particularly convenient because 
only one tunable laser is required. However, the second (ionizing) step generally 
requires high intensity, resulting in power broadening of the first (bound-bound) step. 
Two-photon, two-color detection (e.g., through states from the K(4$s$)+Rb(5$p$) asymptotes) 
may offer some benefits. The ionizing step can be driven with high intensity from a
fixed-frequency pulsed laser, such as a frequency-doubled YAG laser.
If the first step is driven with a narrow-linewidth cw laser at low intensity, rotational
resolution should be achievable.
    
\begin{acknowledgments}
We gratefully acknowledge support from NSF and the University of Connecticut Research Foundation.
We thank Chad Orzel for making the pulsed dye laser available, and Ye Huang and
Hyewon Kim for laboratory assistance.
\end{acknowledgments}

\bibliography{detection}

\end{document}